\documentclass[baaa]{baaa}

\usepackage[pdftex]{hyperref}
\usepackage{subfigure}
\usepackage{natbib}
\usepackage{helvet,soul}
\usepackage[font=small]{caption}

\begin{document}


\journalvol{58}
\journalyear{2015}
\journaleditors{P. Benaglia, D.D. Carpintero, R. Gamen \& M. Lares}


\contriblanguage{1}


\contribtype{1}

\thematicarea{9}

\title{Giant planet formation via pebble accretion}
\subtitle{}


\titlerunning{Giant planet formation via pebble accretion}


\author{O. M. Guilera\inst{1,}\inst{2}}
\authorrunning{O. M. Guilera}


\contact{oguilera@fcaglp.unlp.edu.ar}

\institute{Grupo de Ciencias Planetarias, Instituto de Astrof\'{\i}sica de La Plata (CONICET-UNLP) \and
  Grupo de Ciencias Planetarias, Facultad de Ciencias Astron\'omicas y Geof\'{\i}sicas (UNLP)
}


\resumen{En el marco del modelo cl\'asico de acreci\'on del n\'ucleo, la formaci\'on de un planeta gigante ocurre por dos procesos principales: primero se forma un n\'ucleo masivo por acreci\'on de s\'olidos presentes en el disco protoplanetario; luego, cuando el n\'ucleo excede un valor cr\'{\i}tico (generalmente mayor a $10~\text{M}_{\oplus}$) se dispara la acreci\'on del gas circundante y el planeta acreta grandes cantidades de gas en un per\'{\i}odo corto de tiempo (del orden de $10^5$~a\~nos) hasta que el mismo alcanza su masa final. De esta manera, la formaci\'on de un n\'ucleo masivo tiene que ocurrir cuando a\'un hay gas disponible para ser acretado en el disco. Esto impone una fuerte restricci\'on temporal en la formaci\'on de los planetas gigantes, dado que pr\'acticamente no se observan discos protoplanetarios en estrellas con m\'as de $10^7$~a\~nos. La formaci\'on de n\'ucleos masivos en un tiempo menor a $10^7$~a\~nos por la acreci\'on de planetesimales grandes (con radios $>$ 10 km) solo es posible a partir de discos protoplanetarios masivos. Sin embargo, las tasas de acreci\'on aumentan significativamente para planetesimales de menor tamaño, especialmente para las {\it pebbles}: part\'{\i}culas con tama\~nos del orden del mm y cm, las cuales estan, desde un punto de vista din\'amico, acopladas fuertemente al gas. En este trabajo, analizaremos la formaci\'on de planetas gigantes incorporando las tasas de acreci\'on de {\it pebbles} en nuestro modelo global de formaci\'on planetaria.}

\abstract{In the standard model of core accretion, the formation of giant planets occurs by two main processes: first, a massive core is formed by the accretion of solid material; then, when this core exceeds a critical value (typically greater than $10~\text{M}_{\oplus}$) a gaseous runaway growth is triggered and the planet accretes big quantities of gas in a short period of time until the planet achieves its final mass. Thus, the formation of a massive core has to occur when the nebular gas is still available in the disk. This phenomenon imposes a strong time-scale constraint in giant planet formation due to the fact that the lifetimes of the observed protoplanetary disks are in general lower than 10 Myr. The formation of massive cores before 10 Myr by accretion of big planetesimals (with radii $>$ 10 km) in the oligarchic growth regime is only possible in massive disks. However, planetesimal accretion rates significantly increase for small bodies, especially for pebbles, particles of sizes between mm and cm, which are strongly coupled with the gas. In this work, we study the formation of giant planets incorporating pebble accretion rates in our global model of planet formation.}



\maketitle

\section{Introduction}
\label{sec:intro}

In the standard core accretion model the main question regarding giant planet formation is how to form massive cores before the dissipation of the protoplanetary disk. \citet{Ormel&Klahr2010} and \citet{Lambrechts&Johansen2012} demonstrated that small particles, often called {\it pebbles}, with Stoke number $\text{S}_t \lesssim 1$ are strong coupled to the gas and are very efficiently accreted by the planets. The main difference with planetesimal accretion is that pebbles can be accreted by the full Hill sphere of the planet while planetesimals can only be accreted by a fraction of the Hill sphere, $\alpha^{1/2}~\textrm{R}_{\textrm{H}}$, with $\alpha= \sqrt{\textrm{R}_{\textrm{c}}/\textrm{R}_{\textrm{H}}}$, being $\textrm{R}_{\textrm{c}}$ the core radius of the planet and $\textrm{R}_{\textrm{H}}$ the Hill radius of the planet. The formation of massive cores before 10 Myr by accretion of big planetesimals (with radii $>$ 10 km) in the oligarchic growth regime is only possible in massive disks \citep{Fortier2013,Guilera2014}. Thus, pebble accretion appears as a new alternative in the formation of giant planets \citep{Lambrechts-et-al-2014}. In this work, we study the formation of a massive cores incorporating the pebble accretion rates in our model planet formation \citep{Guilera2010, Guilera2014}.
 
\section{Our model of planet formation}
\label{sec:sec2}

In a series of previous works \citep{Guilera2010, Guilera2011, Guilera2014}. we developed a model which calculates the simultaneous formation of planets immersed in a protoplanetary disk that evolves in time. In this new work, we incorporate some improvements to our previous model, especially the pebble accretion rates given by \citet{Lambrechts-et-al-2014} in order to study the formation of giant planets by pebble accretion. The main characteristics of our model are,

\vspace{0.25cm}

\noindent Planets: 
\begin{itemize}
\item[-] solid cores grow by planetesimal accretion (in the oligarchic regime) or by pebble accretion,
\item[-] gas accretion and the thermodynamic state of the planet envelope are calculated solving the standard equations of stellar evolution.
\end{itemize}

\noindent The protoplanetary disk:
\begin{itemize}
\item[-] the gaseous component evolves as an $\alpha$ accretion disk with photoevaporation,
\item[-] the planetesimal or pebble population evolves by 3 factors: accretion by the planets, migration due to gas drag (3 regimes: Epstein, Stokes and quadratic), and collisional evolution
\end{itemize}

\subsection {Evolution of the disk}
\label{sec:sec2-1}

As we mentioned above, the gas surface density of the disk $\Sigma_g$ evolves as  an  $\alpha$ accretion disk \citep{Pringle1981} with photoevaporation \citep{Dullemond2007}
\begin{eqnarray}
  \frac{\partial \Sigma_g}{\partial t}= \frac{3}{R}\frac{\partial}{\partial R} \left[ R^{1/2} \frac{\partial}{\partial R} \left( \nu \Sigma_g R^{1/2}  \right) \right] + \dot{\Sigma}_w(R), 
\label{eq:eq1}
\end{eqnarray}
where $R$ is the radial coordinate, $\nu$ is the viscosity, and $\dot{\Sigma}_w$ represents the sink term due to photoevaporation. 

Regarding the solid component of the disk, this obeys a continuity equation for the solid surface density $\Sigma_p$
\begin{eqnarray}
  \frac{\partial \Sigma_p}{\partial t} - \frac{1}{R}\frac{\partial}{\partial R} \bigg(Rv_{\text{mig}}(R)\Sigma_p\bigg) = \mathcal{F}(R), 
\label{eq:eq2}
\end{eqnarray}
where $v_{\text{mig}}$ is the planetesimal or pebble migration velocity and $\mathcal{F}$ represents the sink terms due to the accretion by the embryos \citep{Guilera2010}. 

\subsection{Growth of the planets}
\label{sec:sec2-2}

We considered that the cores of the planets grow by planetsimal and pebble accretion. For planetesimals, we use the planetesimal accretion rates given by \citet{Inaba2001}, while for pebbles we use the pebble accretion rates given by \citet{Lambrechts-et-al-2014}. So, the solid accretion rates in our model are given by


\begin{eqnarray}
 \frac{d\textrm{M}_{\textrm{c}}}{dt}= 
  \begin{cases}
    \frac{d\textrm{M}_{\textrm{c}}}{dt}\big|_{\text{planetesimal}}^{\text{Inaba}} & \text{ if $\text{S}_t \ge 1$},  \\
    \\
    \frac{d\textrm{M}_{\textrm{c}}}{dt}\big|_{\text{pebble}}^{\text{L\&J}} & \text{ if $\text{S}_t < 1$},
  \end{cases}
\label{eq:eq3}
\end{eqnarray}
with
\begin{eqnarray}
\frac{d\textrm{M}_{\textrm{c}}}{dt}\big|_{\text{planetesimal}}^{\text{Inaba}}= 2 \textrm{R}_{\textrm{H}}^2\Sigma_p \Omega_{\textrm{P}} \textrm{P}_{\text{coll}}, \text{ if $\text{S}_t \ge 1$}, 
\label{eq:eq4}
\end{eqnarray}
\begin{eqnarray}
  \frac{d\textrm{M}_{\textrm{c}}}{dt}\big|_{\text{pebble}}^{\text{L\&J}} =
  \begin{cases}
    \beta 2 \textrm{R}_{\textrm{H}}^2\Sigma_p \Omega_{\textrm{P}}, \text{ if }  0.1 \le \textrm{S}_t < 1, \\
    \\
    \beta 2 \left( \frac{\textrm{S}_t}{0.1} \right)^{2/3} \textrm{R}_{\textrm{H}}^2\Sigma_p \Omega_{\textrm{P}}, \text{ if }  \textrm{S}_t < 0.1. 
  \end{cases}  
  \label{eq:eq5}
\end{eqnarray}
$\textrm{P}_{\text{coll}}$ is a probability collision \citep[see][for a detail explanation]{Guilera2010}, and $\Omega_{\textrm{P}}$ is the keplerian frequency at the location of the planet. We introduced a factor $\beta$ in the pebble accretion rates. This factor is defined as $\beta= \text{min}(1, \textrm{R}_{\textrm{H}}/\textrm{H}_p)$, and take into account a reduction in the pebble accretion rates if the scale height of small pebbles, $\textrm{H}_p$, could be greater than the Hill radius of the planet. The scale height of the solids is given by \citep{Youdin2007}
\begin{eqnarray}
  \textrm{H}_p= \textrm{H}_g\sqrt{\frac{\alpha}{S_t}}, 
  \label{eq:hp}
\end{eqnarray}
where $\textrm{H}_g$ is the gas disk scale height, and $\alpha$ is the turbulence parameter of the gas disk assuming the \citet{Shakura1973} prescription.

Finally, the gas accretion rate and the thermodynamic state of the planet envelope are calculated solving the standard equations of transport and structure, using an adapted Henyey type code \citep{Fortier2009, Guilera2010}. 

\section{In situ giant planet formation at 5 au}
\label{sec:sec3}

We assumed that the mass of the central star and the mass of the disk are $\text{M}_{\star}= 1~\text{M}_{\odot}$ and $\text{M}_d = 0.05~\text{M}_{\odot}$. The initial gas and solid surface densities are given by 
\begin{eqnarray}
  \Sigma_g &=& \Sigma_g^0 \left( \frac{R}{R_c} \right)^{-\gamma} e^{-(R/Rc)^{2-\gamma}}, \\
  \Sigma_p &=& \eta \Sigma_p^0 \left( \frac{R}{R_c} \right)^{-\gamma} e^{-(R/Rc)^{2-\gamma}}, 
\label{eq:eq6}
\end{eqnarray}
with $R_c= 20$~au and $\gamma= 0.9$ \citep{Andrews2010}. $\eta= 0.25$, if $R < 2.7$~au, or $\eta= 1$, if $R \ge 2.7$~au. The disk is extended beetween 0.1~au and 1000~au using 5000 radial bins logarithmically equally spaced.
 
\begin{figure}[!ht]
  \centering
  \includegraphics[width=0.45\textwidth]{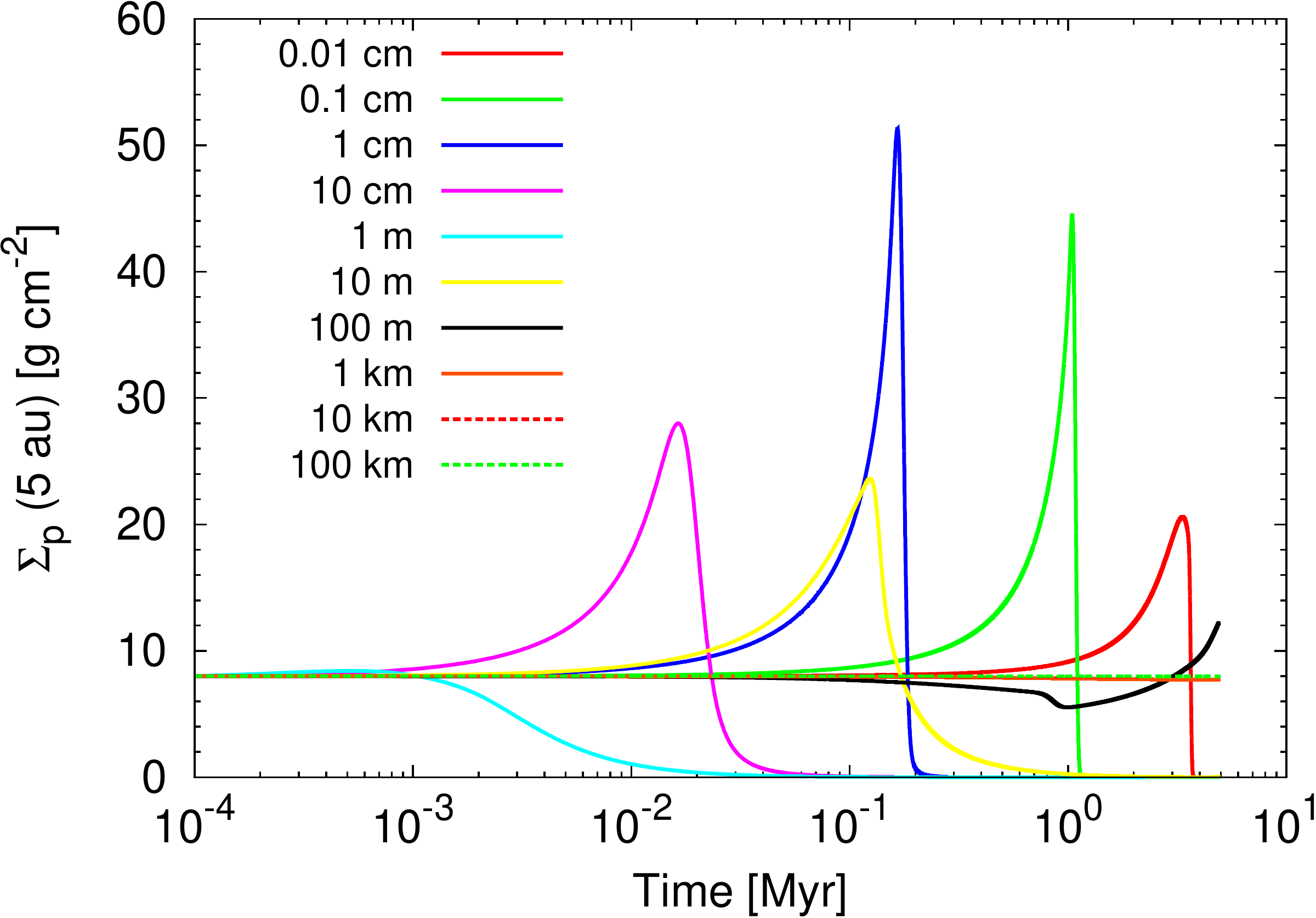}
  \caption{Time evolution of the solid surface density at 5~au. The inward migration of small particles, from the outer region of the disk, significantly increases the surface density. For big planetesimals, the surface density remains almost constant until the dissipation of the disk.}
  \label{fig:fig1}
\end{figure}

Before we calculated the in situ formation of the planet at 5~au, we first analized the evolution of the disk without any planet in it. For simplicity, we considered an unique size for the planetesimals/pebbles along the disk, and we did not consider the collisional evolution of them. So, the solid component of the disk evolves only by planetesimal/pebble migration. Fig~\ref{fig:fig1} shows that the radial drift of small planetesimals and pebbles could play an important role in the formation of massive cores. While the inward migration of material significantly increases the surface density at 5~au, for some sizes (between 1~cm and 10~m) there is a quickly subsequent decline of such surface density. Thus, the planet has to be able to accrete the material before all of it moves inward.  

\begin{figure}[!ht]
  \centering
  \includegraphics[width=0.45\textwidth]{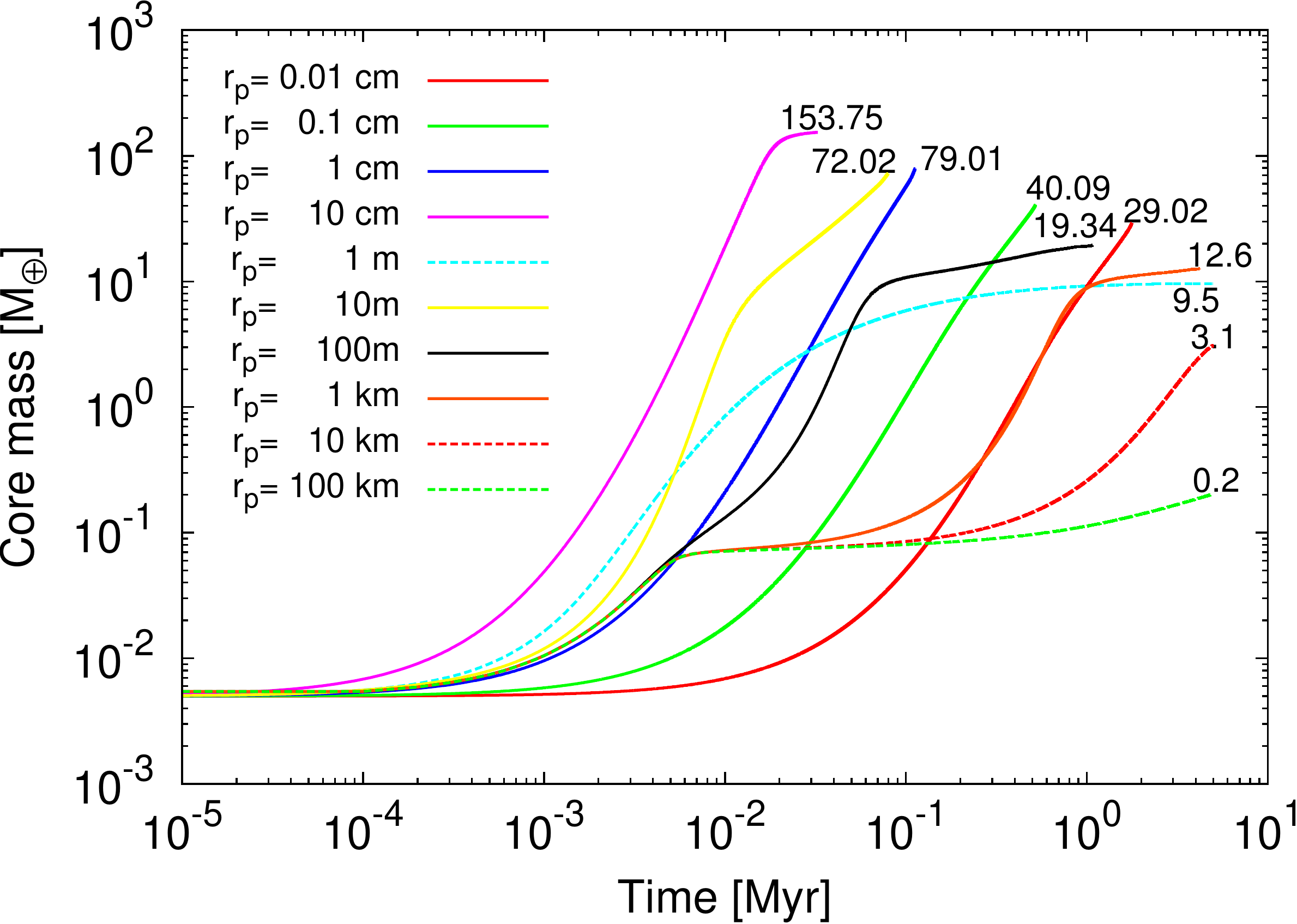}
  \caption{Time evolution of the core mass of the planet. Pebbles ($r_p < 1$~m) are very efficiently accreted and massive cores are formed very quickly. Planetesimals with 1~m $< r_p <$ 100~m are efficiently accreted too, due to the presence of the planet envelope which significantly increases the capture radius of the planet \citep{Guilera2014}. Solid lines represent the cases when the planet achieved the critical mass, and green lines the cases when not.}
  \label{fig:fig2}
\end{figure}

Fig.~\ref{fig:fig2} shows the growth of the planet core as function of time. Simulations stopped when the planet achieved the critical mass (when the envelope mass equaled the core mass) or when the disk was dissipated (at $\sim 5$~Myr). For pebble ($r_p \lesssim 1$~m) and small planetesimales ($1$~m $< r_p < 100$~m) the planet achieved the critical mass very quickly. However, in general the critical core masses are very large. But, \citet{Lambrechts-et-al-2014} showed that when the planet become massive enough ($\text{M}_{core} \gtrsim 20~\text{M}_{\oplus}$), it can perturb the surrounding gas and halts the pebble accretion. 

\begin{figure}[!ht]
  \centering
  \includegraphics[width=0.45\textwidth]{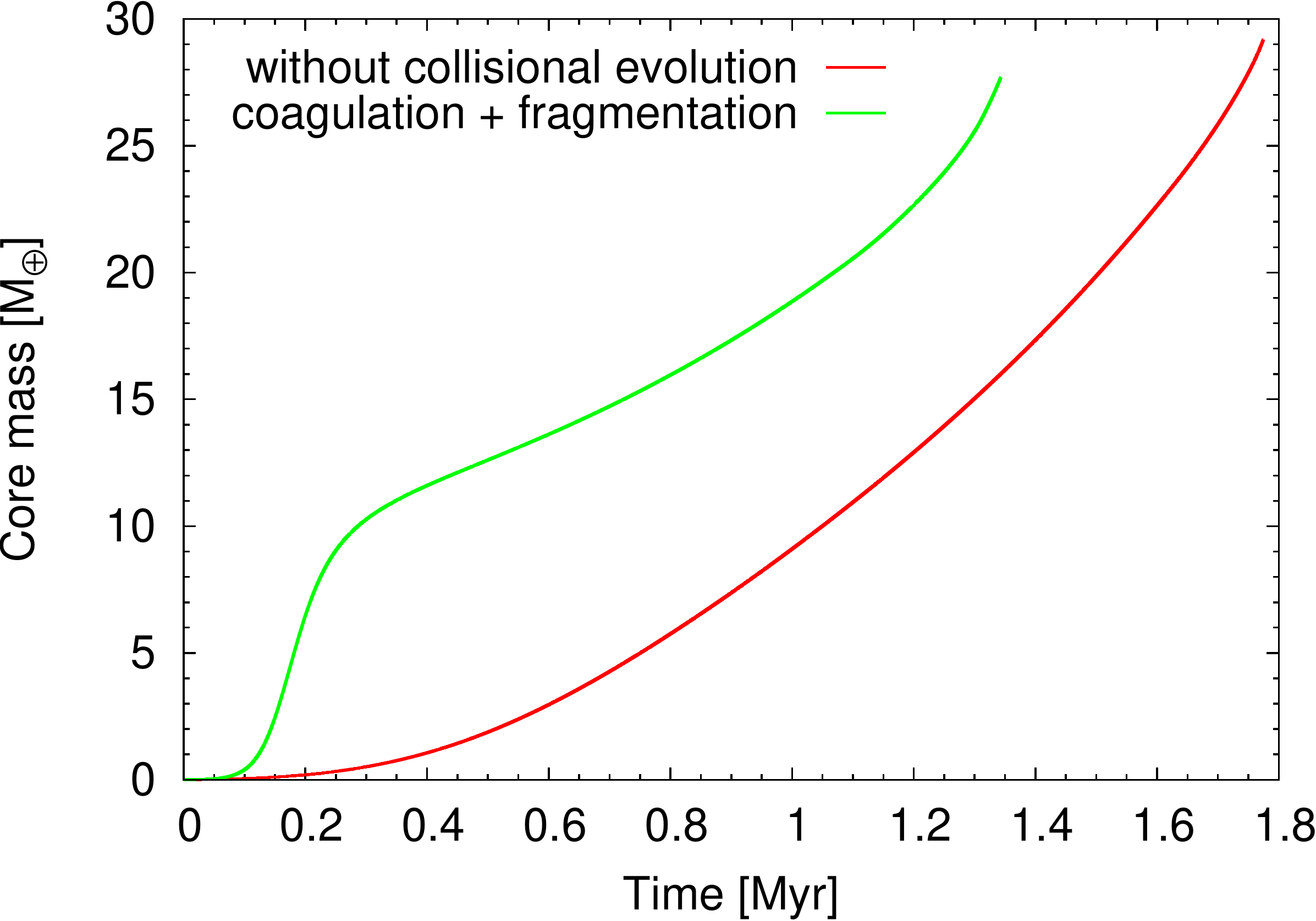}
  \caption{Comparison of the time evolution of the planet core mass between the cases with, and without ($r_p= 0.01$~cm) solid collisional evolution. The coagulation between small pebbles significantly favors the quickly formation of a massive core.}
  \label{fig:fig3}
\end{figure}

Finally, we calculated again the in situ formation of a planet at 5 au, but now considering a planetesimal size distribution. We used 46 size bins between 0.01~cm and 100~km logarithmically equally spaced. Initially, all the solid mass is in the pebbles of 0.01 cm. The collisional evolution of the system is calculated using the model developed in \citet{Guilera2014} (considering coagulation/fragmentation between the particles along the disk). Fig.~\ref{fig:fig3} shows the time evolution of the planet core mass. We can see the incorporation of the collisional evolution of the population of solids allow to the planet reached a massive core of about $10~\text{M}_{\oplus}$ in oly $\sim 0.2$~Myr. 

\section{Conclusions}

Pebble accretion appears as an interesting phenomenon in the formation of giant planets. The high accretion efficiency of these particles could solve the problem of the formation of massive cores before the dissipation of the protoplanetary disk. However, we consider that accurate models of collisional evolution, couple with models of planet formation, are needed due to the fact of the strong dependence between the time-scale of the solid accretion, solid migration, and the sizes of the bodies. Other important question, not treated in this work, is if these pebbles can always reached the core. When the mass of the core is a few times the mass of the Earth, the planet is able to bind a non negligible envelope and pebbles could be destroyed before reached the core. This situation could change the evolution of the growth of the planet, especially the accretion of gas \citep[see][]{Venturini.et.al.2015}.

\begin{acknowledgement}
This work was supported by grants from the Consejo Nacional de Investigaciones Cient\'{\i}ficas y T\'ecnicas and Universidad Nacional de La Plata, Argentina.
\end{acknowledgement}


\bibliographystyle{baaa}
\small
\bibliography{Biblio-aaa2015-Guilera}

\end{document}